\documentclass[preprint,nofootinbib]{article}
\usepackage{epsfig,latexsym,amssymb}
\newcommand{\sect}[1]{\vspace{10pt} \noindent {\bf #1} {\vspace{6pt}}}

\newcommand{\be}{\begin{equation}}
\newcommand{\ee}{\end{equation}}
\newcommand{\bea}{\begin{eqnarray}}
\newcommand{\eea}{\end{eqnarray}}

\begin{document}
\baselineskip=14pt

\begin{center}
{\Large{\bf What if Time Really Exists?}}

\vspace*{0.3in}
Sean M.\ Carroll
\vspace*{0.3in}

\it California Institute of Technology\\ {\tt seancarroll@gmail.com}
\vspace*{0.2in}
\end{center}

\begin{abstract}
Despite the obvious utility of the concept, it has often been argued that time does not exist.  I take the opposite perspective:  let's imagine that time does exist, and the universe is described by a quantum state obeying ordinary time-dependent quantum mechanics.  Reconciling this simple picture with the known facts about our universe turns out to be a non-trivial task, but by taking it seriously we can infer deep facts about the fundamental nature of reality.  The arrow of time finds a plausible
explanation in a ``Heraclitean universe,'' described by a quantum state eternally evolving
in an infinite-dimensional Hilbert space.
\end{abstract}

\sect{Against Time}

Time seems important.  Most of us would have trouble getting through a typical day without looking at a watch, checking a calendar, or anticipating an event scheduled to occur at some specific time.  According to researchers at the {\it Random House Dictionary}, ``time'' is the single most used noun in the English language.  Time is all around us, imposing order on the sequence of happenings in the natural world.

And yet, there is a venerable strain of intellectual history that proclaims that time does not exist.  Partly this claim is pure contrarianism; when something is so obvious and important, declaring that it isn't real is sure to win points for boldness.  But partly, it arises from the difficulty we have in understanding time at a fundamental level.  For something so basic, time is remarkably elusive, and there is a strong temptation to throw up one's hands and proclaim the whole thing is an illusion.  In this essay I will take the opposite tack, doubling down on the contrarian impulse:  time is real, it needs to be dealt with, and by taking the implications of time seriously we are led to important insights concerning the nature of reality.

The denial of time goes back -- well, a long time.  At least to the Presocratics, among whom Parmenides (c. 500 BCE) was famous for proclaiming that change is impossible \cite{parmenides}. A thousand years later, St. Augustine of Hippo reached a similar conclusion, arguing that ``the past is present memory." At the turn of the 20th century, J.M.E. McTaggart presented a proof that time could not exist, as the concept was fundamentally incoherent. These thinkers fall loosely under the rubric of ``presentism'' -- the view that time is an illusion reflecting correlations within an individual state of affairs defined instantaneously, rather than tying together a sequence of such moments. The contemporary mantle of foremost time-denier belongs to Julian Barbour, who has devoted an eloquent book to the subject \cite{Barbour:2000ad}, using language that Parmenides would have found perfectly sympathetic.

In some modern approaches to quantum gravity, presentism has come back with a vengeance.  Quantum mechanics replaces Newton's deterministic laws of motion with Schr\"odinger's deterministic wave equation; in either case, knowledge of the state of the system at any one time is sufficient to determine its future and past evolution in time.  The Schr\"odinger equation can be written (in units where $\hbar=1$)
\be
 \hat H |\Psi\rangle = i \partial_t |\Psi\rangle \,.
  \label{schr}
\ee
Here, $|\Psi\rangle$ is the quantum state, $\hat H$ is an operator called the Hamiltonian, and $\partial_t$ is a derivative with respect to time.  Schr\"odinger's equation uniquely determines the evolution of the quantum state by asserting that the infinitesimal change in the state from one moment to the next is given by acting the Hamiltonian operator on that state.

But in the case of gravity, the Schr\"odinger equation becomes the Wheeler-DeWitt equation:
\be
  \hat H |\Psi\rangle = 0 \,.
  \label{wdw}
\ee
That is to say, allowed states are those for which the Hamiltonian vanishes.  The Wheeler-DeWitt equation embodies a stark form of presentism:  it simply tells us which states the universe can find itself in, and says nothing about any evolution through time. 

However, these perspectives should not convince us to give up on the reality of time.  For one thing, they may not be right; there are deep issues here remaining to be unraveled, both in philosophy and in fundamental physics.   But for another, what is really being questioned here is the idea that time is {\it necessary}, or {\it unique}, or {\it absolute}.  We can think of the universe, according to these arguments, without referring to the concept of time, or without treating it as something special.  But that is very different from saying that time is not {\it useful}, or that it could not be part of a valid description of ultimate reality, or that trying to understand it better would not lead us somewhere interesting.

John Wheeler, following Niels Bohr, liked to admonish physicists to be radically conservative -- to start with a small, reliable set of well-established ideas (conservative), but to push them to their absolute limits (radical) in an effort to understand their consequences.  It is in Wheeler's spirit that I want to ask what the consequences would be if we take time seriously.  What if time exists, and is eternal, and the state of the universe evolves with time obeying something like Schr\"odinger's equation?  This is a point of view that has by no means become obsolete, and may ultimately prove to be indispensable.  We will find that, by taking time seriously, we can conclude a great deal about the deep architecture of reality.

\sect{Lessons of Duality}	

The nature of time is intimately connected with the problem of quantum gravity.  At the classical level, Einstein's general relativity removes time from its absolute Newtonian moorings, but it continues to play an unambiguous role; time is a coordinate on four-dimensional spacetime, and in another guise it measures the spacetime interval traversed by objects moving slower than light.  Quantum mechanics, meanwhile, takes things we formerly considered fundamental, like the position and momentum of a particle, and turns them into mere ``observables'' that imperfectly reflect the reality of the underlying quantum state.  It is therefore perfectly natural to imagine that, in a full theory of quantized gravity, spacetime itself would emerge as an approximation to something deeper.  And if spacetime is an emergent phenomenon, surely time must be.

But the proof of the pudding is in the tasting, and right now the best taste we have of quantum gravity comes from string theory.  Putting aside the (undoubtedly important) question of whether string theory provides the correct theory of everything appropriate to our real world, there is very good evidence that certain formulations of string theory define some perfectly well-defined models of quantum gravity.  And they have crucial implications for the nature of time.

String theory has not solved all of the deep conceptual issues involved in quantizing gravity, but nevertheless has managed to establish firm footholds through a trick known as ``duality." Two theories are said to be dual if they look different, but in fact are isomorphic to each other under some set of transformations.   A famous example is electromagnetism:  if we add magnetic monopoles to Maxwell's equations, we find a theory that is dual to itself under the interchange of electric/magnetic fields and sources.  
But dualities can be much more dramatic than that -- indeed, string theorists have established dualities between theories defined in entirely different numbers of spacetime dimensions.  Such relationships bring to life the idea of {\it holography} -- a theory with gravity, defined in a certain number of dimensions, can actually be equivalent to a theory without gravity, defined in a smaller number of dimensions \cite{'tHooft:1993gx,Susskind:1994vu}.  Holographic duality is a vivid demonstration that a purportedly fundamental concept such as ``the dimensionality of spacetime'' can ultimately be a matter of one's point of view.  

The best example of holography is gauge/gravity duality, pioneered by Juan Maldacena \cite{Maldacena:1997re}.  He discovered that a four-dimensional supersymmetric gauge theory defined on Minkowski space, in the limit of a large number of colors and strong coupling, is equivalent to ten-dimensional supergravity compactified on a five-sphere, with anti-de~Sitter boundary conditions at spatial infinity.  That's a mouthful, but the essential point should be clear:  a full theory of quantum gravity in ten dimensions turns out to be equivalent to an ordinary quantum field theory in four dimensions.  Whatever questions one might have about (this particular version of) quantum gravity must, in principle, have answers in terms of non-gravitational field theory.\footnote{Another formulation of string theory, known as matrix theory, establishes the equivalence of eleven-dimensional supergravity and a non-relativistic theory of zero-dimensional branes \cite{Banks:1996vh}.  Again, a theory of quantum gravity is the same as an ordinary quantum-mechanical theory.}

Duality between a theory with gravity and one without gravity has crucial consequences for our understanding of time.  When quantizing gravity, spacetime itself becomes part of the quantum description, and time seems to disappear according to the Wheeler-DeWitt equation.  But quantization of a non-gravitational theory proceeds straightforwardly in a fixed spacetime background, and time appears as usual in the Schr\"odinger equation.  How are these to be reconciled?

For the purposes of this essay, we can take the lesson of duality to be the following:  whether or not the concept of time is fundamental or unique or necessary, there nevertheless could exist some description of the universe consisting simply of a quantum-mechanical state evolving with time.  If that is true, we have a well-posed question:  can we reconcile what we know about the observable universe with the idea of a wave function evolving with time according to the conventional rules of quantum mechanics?  

This will turn out to be a fairly deep question.  The key insight is to put aside, at least for the moment, all of the complications of spacetime and quantum gravity, and simply think of ordinary non-gravitational quantum mechanics.  Can such a system display the behaviors we see in the real universe?  We can worry about matching on to a spacetime description at the end.

\sect{An Eternal Universe}	

Let's be clear about the scenario we are envisioning to be the case.  Just as in ordinary quantum mechanics, the state of the universe is described by a wave function $|\Psi\rangle$, a ray in a Hilbert space with some number of dimensions.  There exists a Hamiltonian operator $\hat H$ defined on this Hilbert space; we assume that the Hamiltonian is itself independent of time.  The wave function evolves with time according to the Schr\"odinger equation,
\be
 \hat H |\Psi\rangle = i \partial_t |\Psi\rangle \,.
  \label{schr2}
\ee
We are going to be radically conservative, and ask whether this prosaic setup can possibly describe our universe.  And we are going to imagine that the time parameter $t$ corresponds more or less to some version of time as we experience it.  What we are not worrying about, for the moment, is what that wave function means -- its interpretation in terms of things we observe around us in the world.  None of these assumptions is beyond questioning -- quantum mechanics may be incomplete, time may only be emergent in a semi-classical description, or the Hamiltonian may be time-dependent -- but it is worth our effort\footnote{One is tempted to say, ``worth our time.''  But one resists.} to pursue their ramifications and see where we end up.

Assuming the validity of the Schr\"odinger equation has a deep, if somewhat obvious, consequence:  time stretches for all of eternity.  In classical mechanics, singularities in phase space can disrupt the evolution, causing time to grind to a halt. But in quantum mechanics, unitary evolution ensures that there there is no boundary to time; the variable $t$ runs from $-\infty$ to $+\infty$.  The modern idea that time does have a beginning arises from the existence of a Big Bang singularity in cosmological models based on general relativity.  But from our current perspective, that is an outmoded relic of our stubborn insistence to think in terms of spacetime, rather than directly in terms of the quantum state.  Classical general relativity, after all, is not correct; at some point it must be subsumed into a quantum description of gravity.  We therefore imagine that the classical Big Bang corresponds to some particular kind of quantum state, which may be obscure from the perspective of our current knowledge, but will ultimately be resolved.  It follows, under our assumptions, that there was something before the Big Bang, and time stretches back into the infinite past.

But this raises a problem, well known to anyone who has thought about the mysteries of time:  within our observable universe, time has an arrow.  Entropy was small near the Big Bang, is somewhat larger today, and will be even larger in the future.  On macroscopic scales where the concept of entropy makes sense, the world around us is characterized by {\it irreversible} processes, from the mixing of milk into coffee to the collapse of a star to form a black hole.  But the Schr\"odinger equation, on which this is all purportedly based, is perfectly reversible.  The Hamiltonian operator may not respect time-reversal invariance (as in the weak interactions of the Standard Model), but that is beside the point.  As long as the the Schr\"odinger equation holds, quantum time evolution is perfectly unitary (information-preserving) -- given the current quantum state, we can reliably reconstruct the past just as well as the future.  How can we reconcile this with macroscopic irreversibility?

Ludwig Boltzmann taught us in the 1870's that entropy is a measure of the volume of the state space corresponding to macroscopically indistinguishable microstates.  If a state is macroscopically unique or nearly so, it has a low entropy; if a state is one of a very large number that look the same from a macroscopic (coarse-grained) point of view, it has a high entropy.  From that point of view, it might seem like no surprise that entropy tends to increase; starting in a low-entropy state, generic evolution through time is overwhelmingly likely to move the system toward higher entropy, for the simple reason that there are overwhelmingly more high-entropy states than low-entropy ones.

But that explanation begs a crucial question:  why was the entropy ever low to begin with?  This question ultimately belongs in the realm of cosmology.  We see entropy increasing all around us, in our kitchen when we break an egg to make an omelet.  But the reason why we ever had access to a low-entropy configuration such as an egg is ultimately because the environment of the Earth is a low-entropy place, which is because the Solar System arose from an even lower-entropy protostellar cloud, which evolved out of the even lower-entropy primordial plasma, which arose from a yet lower-entropy configuration near the Big Bang.  

Boltzmann's insight explains why, beginning with a low-entropy state, irreversible processes that increase the entropy are the most natural things in the world.  But the fact that we ever began with a low-entropy is not natural at all.  If time had a beginning, we could at least imagine that the beginning was simply characterized by special low-entropy boundary conditions.  But if time is eternal, that option is closed to us.  Can the observed arrow of time be explained by the apparently reversible physics embodied in the Schr\"odinger equation?

\sect{The Universe as a Box of Gas?}

Boltzmann was well aware of the problem of initial conditions that grew out of his statistical understanding of entropy, and he proposed a number of possible resolutions.  One of these was that time had a beginning, with special boundary conditions.  But he also considered the possibility that time was eternal.  In that case, assuming that the space of states of the universe is finite, low-entropy conditions were guaranteed to happen eventually, simply as statistical fluctuations.  The universe would be like a box of gas, left to itself for all eternity:  usually it would be in thermal equilibrium, but there would be occasional random fluctuations that arranged its components in a state of very low entropy \cite{boltzmannvzermelo2}.
\begin{quote}
There must then be in the universe, which is in thermal equilibrium as a whole and therefore dead, here and there relatively small regions of the size of our galaxy (which we call worlds), which during the relatively short time of eons deviate significantly from thermal equilibrium.  Among these worlds the state probability increases as often as it decreases.  
\end{quote}
This is a remarkable quote, which anticipates a number of ideas in contemporary cosmology.  Boltzmann understands that a typical state of a system should be one of high entropy, thermal equilibrium.  But in thermal equilibrium, life cannot exist, for the simple reason that nothing ever happens.  So Boltzmann invokes the anthropic principle: life will only be found in those regions of the universe that are hospitable to the existence of life -- in this case, regions with an entropy gradient.  And, due to statistical fluctuations, such regions will appear here and there in space and time, creating what we would now call a ``multiverse.''  Boltzmann suggests that the universe we see around us is part of the aftermath of a substantial downward fluctuation in entropy, purely statistical in nature.

In fact he was not the first to suggest such a scenario, by a wide margin.  In the first century BCE, the Latin philosopher Lucretius meditated on the origin of things.  He was an atomist, in the tradition of Democritus and Epicurus, and proposed an explanation for how the orderly universe around us could arise from the mindless motion of individual atoms \cite{lucretius}.
\begin{quote}
For surely the atoms did not hold council, assigning order to each, flexing 
their keen minds with questions of place and motion and who goes where.  But shuffled and jumbled in many ways, in the course of endless time they are buffeted, driven along, chancing upon all motions, combinations.  At last they fall into such an arrangement as would create this universeÉ
\end{quote}
This is another remarkable quote, anticipating the core of Boltzmann's idea with great fidelity, without the benefit of anything we would recognize as modern physics.

Sadly, this scenario, in which the low entropy of our early universe is explained as a random statistical fluctuation in a finite eternal universe, does not work.  The reason was explained by yet another deep thinker, Sir Arthur Eddington, in yet another remarkable quote, this one from 1931 \cite{eddington}.
If the universe is a statistical fluctuation within an equilibrium ensemble, Eddington argues,
\begin{quote}
A universe containing mathematical physicists will at any assigned date be in the state of maximum disorganization which is not inconsistent with the existence of such creatures.
\end{quote}
Eddington recognized that Boltzmann's microscopic description of entropy did not only allow for statistical fluctuations, it quantified precisely how often they would occur.  The probability of such a fluctuation is proportional to $e^{-\Delta S}$, where $\Delta S$ is the decrease in entropy.  Larger fluctuations in entropy are much more rare.  The idea that the universe is usually in thermal equilibrium, with occasional random excursions into low-entropy configurations, therefore makes a very clear prediction:  given the constraints of the anthropic principle, we should find ourselves in a \emph{minimum possible} deviation from maximum entropy.  For Eddington, that meant a single mathematical physicist, suddenly assembled as a random fluctuation out of the surrounding thermal radiation.  In fact we can do away with the body of the physicist, and distill it down to a disembodied ``Boltzmann Brain'' that exists only long enough to appreciate that the rest of the universe is in thermal equilibrium \cite{Albrecht:2004ke}. The Boltzmann-Lucretius universe predicts strongly that the vast majority of observers are of this type.

But this is a prediction which, manifestly, has been falsified.  It has been suggested that a careful application of Bayesian analysis makes this problem go away -- we shouldn't ask what happens to a ``typical observer,'' we should ask what happens to observers like ourselves \cite{Hartle:2007zv}.  But that doesn't actually extract us from this particular dilemma.  In the thermal-fluctuation scenario, given \emph{any} set of observations (and records and memories) of any one particular observer, it is overwhelmingly likely that everything else in the universe is in thermal equilibrium.  This is clearly not the universe in which we live.

The Boltzmann-Lucretius model is therefore incorrect.  But it didn't assume very much:  only that the universe is eternal, obeys time-independent and information-preserving dynamical laws, and features a state space of finite extent.  Quantum mechanics hadn't been invented at the time, but all of these features are characteristic of conventional quantum mechanics defined on a finite-dimensional Hilbert space.  So the problems of this model are not simply a matter of academic or historical curiosity; they represent severe problems for any theory of the universe founded on these 
principles.\footnote{Dyson, Kleban, and Susskind \cite{Dyson:2002pf} have argued that
de~Sitter space, which describes a universe dominated by a positive cosmological constant,
requires a local description in each Hubble volume, described by a finite-dimensional
Hilbert space.  They recognized the phenomenological problems with such a scenario,
which spurred much of the contemporary interest in Boltzmann Brains.}

\sect{The Neverending Story}

\begin{figure}
\begin{center}
\includegraphics[width=0.45\textwidth]{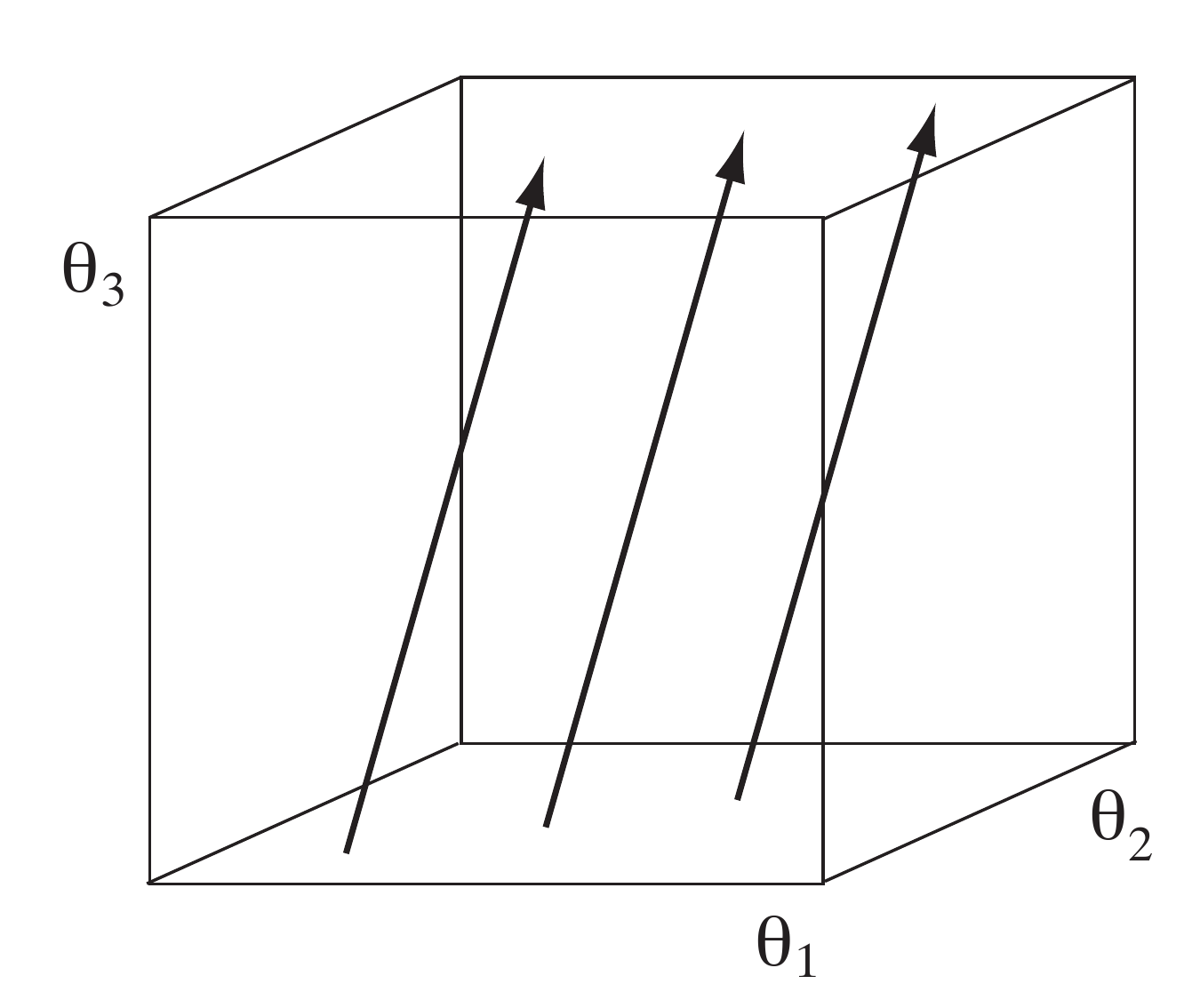}
\end{center}
\caption{Torus of phases.}
\label{torus}
\end{figure}

The time evolution of a state in quantum mechanics is extremely simple.  In terms of
energy eigenstates $\hat H|\phi_a\rangle = E_a |\phi_a\rangle$, we can expand an
arbitrary state as
\be
  |\Psi(t)\rangle = \sum_a e^{-iE_a t}\psi_a |\phi_a(0)\rangle\,.
\ee
Note that the basis states $|\phi_a(0)\rangle$ and the coefficients $\psi_a$ are
completely time-independent; all of the time evolution is encoded in the phases $e^{-iE_a t}$.
In contrast to the arbitrarily complicated evolution of a (non-integrable) classical system, all a 
quantum state ever does is move in circles.  As shown in Figure~\ref{torus}, the
component corresponding to each energy eigenstate rotates in phase $\theta_a$ with a frequency
given by the eigenvalue $E_a$; together, these phases form a torus on which
quantum evolution takes place.

In the real world, the reason why quantum evolution often seems complicated is because we
don't actually know what the energy eigenstates are, in terms of easily observable 
quantities.  Instead, we work in terms of approximate eigenstates, for which the evolution can
be much more complicated.  But if we are thinking about what can possibly happen over the lifetime
of an eternal universe, straight-line motion in the torus of phases captures the entire
story.  In particular, when the phases line up in the right way, the state fluctuates into a
low-entropy configuration (from the coarse-grained point of view), as in the
Boltzmann-Lucretius scenario.  Indeed, the evolution will return arbitrarily closely to any
specific point in its development an infinite number of times, in accordance with the 
Poincar\'e recurrence theorem (and brining to life Friedrich Nietzsche's image of 
eternal return).

The only way to avoid the phenomenological problems associated with the 
Boltzmann-Lucretius scenario is to prevent such recurrences, and the only way to do that
(within our assumptions) is to stretch the time between recurrences to infinity.  In a 
quantum system with a discrete Hilbert space, the time scale associated with the
subspace spanned by energy eigenstates $|\phi_a\rangle$ and $|\phi_b\rangle$ is
\be
  \tau_{ab} = \frac{2\pi}{|E_a - E_b|}\,. 
\ee
If the set of energy eigenvalues is finite, there will be a maximum possible value of
$\tau_{ab}$, and recurrences are inevitable.  
But imagine that the Hilbert space is infinite-dimensional, and furthermore that 
the energy eigenvalues have
an accumulation point: a value $E_*$ such that there are an infinite number of 
eigenvalues within any finite distance of $E_*$.  In that case, the recurrence time
gets pushed off to infinity.  There will be quantum states whose evolution never
returns to any previous configuration -- novel things keep happening, infinitely
far into the future.  We can refer to such a situation as a ``Heraclitean'' universe, 
after Parmenides's fellow Presocratic philosopher Heraclitus, who famously insisted on the
primacy of change and flux:  one cannot step into the same river twice.\footnote{We are
far from doing justice to Heraclitus's actual philosophy, but his name serves as
a useful label.}

Our conclusion that the Hilbert space of the universe needs to be infinite-dimensional
might not seem very startling; the universe is a big place, why should we be surprised
that it requires a big Hilbert space?  What is remarkable is how little input was
required to deduce this fundamental fact about the theory of everything.
When we increase entropy by breaking an egg, we are taking advantage of the
entropy gradient characterizing the evolution of our observable universe.  But the
combination of such a departure from equilibrium with the assumptions of ordinary
quantum evolution in a finite-dimensional Hilbert space leads to a robust prediction:
every new experiment we perform should reveal that the rest of the universe actually
is in equilibrium.  Because this turns out not to be true, either the universe violates
the rules of conventional time evolution in quantum mechanics, or there must be
an accumulation point of energy eigenvalues in an infinite-dimensional Hilbert
space.  Either possibility is perfectly plausible, but the latter seems like a less
dramatic leap into speculative physics, and at the very least is worth taking
seriously.

\sect{Worlds Without End}

So far we have stuck to our program of ignoring specific issues about spacetime and
cosmology, to focus on the structure of a quantum-mechanical model that would be
consistent with the Heraclitean property of non-recurrent change throughout all of
eternity.  In such a setup, it is not surprising to witness an arrow of time; the entropy can
increase because entropy can always increase, as there is no such thing
as an equilibrium state.

Nevertheless, there is a long distance between insisting on an accumulation point
of energy eigenvalues and actually presenting a compelling physical model of the
evolution of the universe.  At some point, a spacetime interpretation is going to be necessary.
Unfortunately, the current state of the art is not up to the task; but nothing stops us from
making some informed guesses.

A few years ago, Jennifer Chen and I considered the question of what would 
constitute a ``natural,'' high-entropy state of the universe \cite{Carroll:2004pn}.
Without a complete theory of quantum gravity, and a corresponding understanding of
how microstates get coarse-grained into macrostates, we cannot reliably calculate
the entropy of the universe.  But we do know that it has been increasing, and we have
every expectation that it will continue to do so.  As our universe expands, it is
increasingly dominated by vacuum energy.  Currently, structures are still forming
and complex life forms are riding the wave of entropy generated by hot suns shining
in cold skies.  But ultimately those stars will grow dim, galaxies will collapse into
black holes, black holes will evaporate, and all we will 
be left with is an increasingly thin gruel of elementary particles in a background
of vacuum energy.  That, then, is a high-entropy state:  a nearly-empty universe 
suffused with a tiny amount of vacuum energy.

How can that be reconciled with eternal evolution in a Heraclitean cosmology?
An obvious answer is that de~Sitter space, the solution to Einstein's equation in the
presence of a positive cosmological constant, is unstable; there must be some way
for it to undergo a transition into a state with even more entropy.  Chen and I imagined that
the mechanism was the quantum creation of baby universes, as suggested
by Farhi, Guth, and Guven \cite{Farhi:1989yr}; a closely related scenario is the
``recycling universe'' idea of Garriga and Vilenkin, in which an entire Hubble patch
of universe spontaneously jumps to a larger value of the vacuum energy \cite{Garriga:1997ef}.

We are a long way from understanding the details of such pictures, or indeed whether
they make much physical sense at all.  But there are robust features of the models
that may very well survive as part of a better-developed understanding.  The crucial
point is that the universe finds itself in a state that will never settle down into
equilibrium.    In such a situation, the kind of entropy gradient we currently find ourselves
in is perfectly natural; entropy is growing because entropy can always grow.

In fact, entropy can grow both into the far future and into the far past; the overall multiverse
can be completely symmetric with respect to time.  Think of two particles moving on
straight lines in an otherwise empty three-dimensional space.  No matter how we choose
the lines, there will always be some point of closest approach, while the distance between
the particles will grow without bound sufficiently far in the future and the past.  We are
suggesting that the evolution of entropy in the universe takes a similar form.  
This is a very different scenario from the various forms of eternal cosmologies that
feature a low-entropy ``bounce'' that replaces the Big Bang
\cite{bounces}; in the
picture advocated here, while there is some moment of minimum entropy in the history
of the universe, that need not be an especially low entropy, any more than the
closest points on two lines in three dimensions need be especially close.

Clearly, a lot more work needs to be done to bring the idea of a Heraclitean cosmology
in line with specific scenarios about the evolution of spacetime and the particular
features of our local universe.  But we are led to consider it by a very mundane
goal -- understanding the arrow of time that is so crucial to our everyday lives -- and
a very reasonable, if far from unimpeachable, set of assumptions -- a quantum state
evolving in time according to the conventional Schr\"odinger equation with a
time-independent Hamiltonian.  From that starting point, we are led to conclude that
the Hilbert space must be infinite-dimensional, with at least one accumulation point for the
set of energy eigenvalues, and to the suggestions that the de~Sitter phase toward which our
current universe is evolving is somehow an unstable configuration, and that the very
far past of our universe could be experiencing an arrow of time directed in the
opposite sense to our own.  That is a great deal of output from a rather small amount
of input.

In Parmenides's work {\it On Nature}, the truth is revealed through the words of a goddess.  
We have no such luxury, so have to work for ourselves at observing the universe and
applying the scientific method.  But we may be able to see just as far.


\begin{thebibliography}{99}

\bibitem{parmenides}
D.~Gallop, {\it Parmenides of Elea},
University of Toronto Press (1991).

%\cite{Barbour:2000ad}
\bibitem{Barbour:2000ad}
  J.~Barbour,
  {\it The End Of Time: The Next Revolution In Physics,}
%\href{http://www.slac.stanford.edu/spires/find/hep/www?irn=4631080}{SPIRES entry}
Oxford Univ. Pr. (2000).

%\cite{'tHooft:1993gx}
\bibitem{'tHooft:1993gx}
  G.~'t Hooft,
  %``Dimensional reduction in quantum gravity,''
  arXiv:gr-qc/9310026.
  %%CITATION = GR-QC/9310026;%%

%\cite{Susskind:1994vu}
\bibitem{Susskind:1994vu}
  L.~Susskind,
  %``The World As A Hologram,''
  J.\ Math.\ Phys.\  {\bf 36}, 6377 (1995)
  [arXiv:hep-th/9409089].
  %%CITATION = JMAPA,36,6377;%%

%\cite{Maldacena:1997re}
\bibitem{Maldacena:1997re}
  J.~M.~Maldacena,
  %``The large N limit of superconformal field theories and supergravity,''
  Adv.\ Theor.\ Math.\ Phys.\  {\bf 2}, 231 (1998)
  [Int.\ J.\ Theor.\ Phys.\  {\bf 38}, 1113 (1999)]
  [arXiv:hep-th/9711200].
  %%CITATION = IJTPB,38,1113;%%
  
%\cite{Banks:1996vh}
\bibitem{Banks:1996vh}
  T.~Banks, W.~Fischler, S.~H.~Shenker and L.~Susskind,
 % ``M theory as a matrix model: A conjecture,''
  Phys.\ Rev.\  D {\bf 55}, 5112 (1997)
  [arXiv:hep-th/9610043].
  %%CITATION = PHRVA,D55,5112;%%

\bibitem{boltzmannvzermelo2}
L.\ Boltzmann, Annalen der Physik {\bf 60}, 392 (1897);  
trans. in {\it Kinetic Theory}, ed. S.~G.~Brush
(Oxford, 1966), p. 412.

\bibitem{lucretius}
Lucretius, {\it On the Nature of Things (De rerum natura)}, edited and translated by Anthony M. Esolen (Baltimore: Johns Hopkins University Press, 1995).

\bibitem{eddington} 
A.~S.~Eddington, {\it Nature} {\bf 127}, 3203 (1931);
reprinted in {\it The Book of the Cosmos: Imagining the Universe from
Heraclitus to Hawking}, ed. D.~R.~Danielson (Perseus,
Cambridge, Mass., 2000), p.\ 406.

%\cite{Albrecht:2004ke}
\bibitem{Albrecht:2004ke}
  A.~Albrecht and L.~Sorbo,
  %``Can the universe afford inflation?,''
  Phys.\ Rev.\  D {\bf 70}, 063528 (2004)
  [arXiv:hep-th/0405270].
  %%CITATION = PHRVA,D70,063528;%%
  
%\cite{Hartle:2007zv}
\bibitem{Hartle:2007zv}
  J.~B.~Hartle and M.~Srednicki,
  %``Are We Typical?,''
  Phys.\ Rev.\  D {\bf 75}, 123523 (2007)
  [arXiv:0704.2630 [hep-th]].
  %%CITATION = PHRVA,D75,123523;%%

%\cite{Dyson:2002pf}
\bibitem{Dyson:2002pf}
  L.~Dyson, M.~Kleban and L.~Susskind,
  %``Disturbing implications of a cosmological constant,''
  JHEP {\bf 0210}, 011 (2002)
  [arXiv:hep-th/0208013].
  %%CITATION = JHEPA,0210,011;%%

%\cite{Carroll:2004pn}
\bibitem{Carroll:2004pn}
  S.~M.~Carroll and J.~Chen,
  %``Spontaneous inflation and the origin of the arrow of time,''
  arXiv:hep-th/0410270.
  %%CITATION = HEP-TH/0410270;%%
  
  %\cite{Farhi:1989yr}
\bibitem{Farhi:1989yr}
  E.~Farhi, A.~H.~Guth and J.~Guven,
  %``IS IT POSSIBLE TO CREATE A UNIVERSE IN THE LABORATORY BY QUANTUM
  %TUNNELING?,''
  Nucl.\ Phys.\  B {\bf 339}, 417 (1990).
  %%CITATION = NUPHA,B339,417;%%

%\cite{Garriga:1997ef}
\bibitem{Garriga:1997ef}
  J.~Garriga and A.~Vilenkin,
  %``Recycling universe,''
  Phys.\ Rev.\  D {\bf 57}, 2230 (1998)
  [arXiv:astro-ph/9707292].
  %%CITATION = PHRVA,D57,2230;%%

%\cite{bounces}
\bibitem{Gasperini:1992em}
  M.~Gasperini and G.~Veneziano,
  %``Pre - big bang in string cosmology,''
  Astropart.\ Phys.\  {\bf 1}, 317 (1993)
  [arXiv:hep-th/9211021];
  %%CITATION = APHYE,1,317;%%
  J.~Khoury, B.~A.~Ovrut, P.~J.~Steinhardt and N.~Turok,
  %``The ekpyrotic universe: Colliding branes and the origin of the hot big
  %bang,''
  Phys.\ Rev.\  D {\bf 64}, 123522 (2001)
  [arXiv:hep-th/0103239];
  %%CITATION = PHRVA,D64,123522;%%
  %\cite{Bojowald:2006da}
  M.~Bojowald,
  %``Loop quantum cosmology,''
  Living Rev.\ Rel.\  {\bf 8}, 11 (2005)
  [arXiv:gr-qc/0601085].
  %%CITATION = 00222,8,11;%%






\end{thebibliography}
\end{document}